\documentclass{PoS}

\usepackage{slashed}
\newcommand\Dslash{\slashed D}
\newcommand\Tr{\mathop{\rm Tr}}
\makeatletter
\@ifundefined{inlinecite}
{\newcommand\inlinecite{\cite}}{}
\@ifundefined{texorpdfstring}
{}{}
\@ifundefined{unichar}
{\newcommand\myunichardef[3]{\expandafter\providecommand\csname text#1\endcsname
                            {#2}}}
{\newcommand\myunichardef[3]{\expandafter\providecommand\csname text#1\endcsname
                            {\unichar{"#3}}}}
\makeatother
\myunichardef{composetilde}{\string~}{0303}
\myunichardef{composemacron}{bar}{0305}
\myunichardef{subscriptthree}{\string_3}{2083}

\newcommand\myIm{\mathop{{\cal I}\!\!m}}
\newcommand\myRe{\mathop{{\cal R}\!e}}

\newcommand\doi[1]{\href{http://dx.doi.org/#1}{doi:#1}}
\newcommand\eprint[2][]{\href{http://arXiv.org/abs/#2}{arXiv:#2\relax
                        \ifx\relax#1\relax\else{ }[#1]\fi}}

\newbox\dpbox
\setbox\dpbox\hbox{}
\dp\dpbox=4\jot

\title{Neutron Electric Dipole Moment from\\ quark Chromoelectric
       Dipole Moment}

\ShortTitle{nEDM from qCEDM}

\author{\speaker{Tanmoy Bhattacharya}\\
        Los Alamos National Laboratory and Santa Fe Institute\\
        E-mail: \email{tanmoy@lanl.gov}}

\author{Vincenzo Cirigliano, Rajan Gupta, Emanuele Mereghetti, Boram Yoon\\
        Los Alamos National Laboratory\\
        E-mail: \email{cirigliano@lanl.gov},
                \email{rg@lanl.gov},
                \email{emereghetti@lanl.gov},
                \email{boram@lanl.gov}}

\abstract{The connection between a regularization-independent symmetric momentum substraction (RI-$\tilde{\rm S}$MOM) and the $\overline{\rm MS}$ scheme for the quark chromo EDM operators is discussed.  A method for evaluating the neutron EDM from quark chromoEDM is described. A preliminary study of the signal in the matrix element using clover quarks on a highly improved staggered quark (HISQ) ensemble is shown.}

\FullConference{The 33rd International Symposium on Lattice Field Theory\\
		14--18 July 2015\\
		Kobe International Conference Center, Kobe, Japan*}

\begin{document}

\section{Introduction}

A very stringent constraint on physics beyond the standard model (BSM) comes from bounds on the electric dipole moments (EDM) of elementary particles. The observed baryon density in our universe, \(n_B/n_\gamma = 6.1^{+0.3}_{-0.2} \times 10^{-10}\)~\cite{WMAP+COBE}, is far larger than the expected freeze-out ratio of \(n_B/n_\gamma \approx 10^{-20}\)~\cite{KolbTurner}, and is due to a baryon-antibaryon asymmetry.  Since standard inflationary models of cosmology cannot accommodate such a large asymmetry as an initial condition, net baryon number needs to have been generated during the post-inflationary evolution of the universe. Sakharov~\cite{Sakharov} stated the minimal conditions on particle physics models for such baryogenesis to be possible, and efforts to design cosmological baryogenesis models without these conditions have not been successful~\cite{nonSakharov}. These Sakharov conditions are (i) baryon number violation, (ii) violations of both the charge-conjugation (C) and combined charge-conjugation-parity (CP) symmetries, and (iii) out of equilibrium evolution. Though the standard model of particle physics has all the required ingredients, the CP violation is too small to produce the observed baryon asymmetry, and one needs to look beyond the standard model for new sources of CP asymmetry.

Most sources of CP violation produce electric dipole moments for particles with non-zero spin.  The standard model has two sources of CP violation: (i) the complex phase in Cabibo-Kobayashi-Maskawa (CKM) quark mixing matrix and (ii) a possible CP-violating mass term or an effective \(\Theta G \tilde G\) gluonic interaction related to QCD instantons.  The latter is ineffective in solving the baryogenesis problem since instanton effects are suppressed at high temperatures relevant for early universe, but is also known to be very small from limits on neutron EDM (nEDM), \(\Theta \lesssim 10^{-10}\)~\cite{Crewther}.  The former gives rise to a tiny contribution to nEDM, \(\sim 10^{-32}\,\hbox{e-cm}\)~\cite{Dar}, much below the experimental limits.  It is, therefore, interesting to attempt to constrain sources of BSM CP violation using experimental constraints on nEDM.

A simple way to parameterize the effect on experimental measurements of BSM physics is within the effective field theory (EFT) framework.  In this approach, we evaluate the low energy effects of BSM physics in terms of higher-dimension operators suppressed by powers of the BSM mass scale \(M_{\rm BSM}\). These operators can be enumerated purely on the basis of their symmetry properties and their dimension\footnote{Due to asymptotic freedom in the standard model, the anomalous dimensions are small, and one can use the engineering dimension in this power counting.} without reference to any particular choice of BSM theory. The low energy hadronic analysis is performed treating these operators as a perturbation to the standard model action, and particular choices for BSM physics, along with renormalization group flow, dictate only the coefficients of these operators. A detailed discussion of experimentally accessible quantities and how they constrain the various operators in an effective field theory analysis of the CP violating sector of BSM physics is given in Ref.~\inlinecite{PospelovRitz}.

\subsection{Enumeration of Operators}

The standard model CP violation occurs primarily in the Higgs-Yukawa sector.  We will work below the scale of the Higgs and electroweak sector, so this CP violation could also be included in the EFT framework that we use below.  Though these effects are suppressed only by the weak vacuum expectation value, \(v_{\rm EW} \approx 246~\rm GeV\), instead of the larger \(M_{\rm BSM}\sim 1~\rm TeV\), flavor mixing and chiral suppression in the standard model makes them very small, and we will, therefore, neglect them in further discussions. 

At dimensions 3 and 4, the only operators that violate CP are the CP violating mass terms {\(\bar\psi \gamma_5 \tau \psi\)} (we use \(\tau\) to represent a general flavor structure), and the topological charge {\(G_{\mu\nu}\tilde G^{\mu\nu}\)}.  We discuss later that because of the freedom to choose the phases of the fermion fields, one can remove all except one of these operators. The remaining operator is anomalously small in the standard model, and one often invokes a Peccei-Quinn mechanism~\cite{PecceiQuinn} to explain it.

Because of the chiral structure of the standard model, the operators at dimension 5 are suppressed by \(v_{\rm EW}/M^2_{\rm BSM}\). These are the electric dipole moments {\(\bar\psi\Sigma_{\mu\nu}\tilde F^{\mu\nu}\tau\psi\)} of the quarks and charged leptons and the chromo-electric dipole moments (cEDM) {\(\bar\psi\Sigma_{\mu\nu}\tilde G^{\mu\nu}\tau\psi\)} of the quarks.  These are the only operators we consider here; in particular BSM theories, however, the effects of the dimension 6 operators that are suppressed by \(1/M^2_{\rm BSM}\), which are the Weinberg operator (also called the gluon chromo-electric moment) \(G_{\mu\nu}G_{\lambda\nu}\tilde G_{\mu\lambda}\), and a set of four-Fermi operators, may not be much smaller.

\section{Renormalization and Mixing}

Before we consider the evaluation of these matrix elements, let us consider the phase choice of fermions and determine the invariant combinations that control the physical CP violation.  Towards this, we note that CP and chiral symmetry do not commute: a chiral rotation \(\chi\) provides an outer automorphism for the \(CP\) symmetry group: if \({\rm CP}\) is the operator implementing the group transformation,  the operator \({\rm  CP}_\chi \equiv \chi^{-1}{\rm CP}\chi\) implements an inequivalent \(CP\) transformation that differs only by a choice of phases.  Under the usual choice of phases, the left and right handed fermion fields \(\psi_{L,R}\) transform into each other with the same phase, {\it i.e.,} 
\(
\psi_{L,R}^{CP} = i\gamma_4 C \bar\psi^T_{L,R}\,,\label{eq:std}
\)
where \(\bar\psi^T_{L,R}\equiv ({\psi_{R,L}}^\dagger\gamma_0)^T\) and \(C\) depends on the representation of the gamma matrices.  Under the chiral rotation, however, the left and right chiral fields pick up opposite phases:  \(\psi_{L,R}^\chi = e^{\pm i \chi}\psi_{L,R}\).  As a result, the \(CP_\chi\) transformation is distinguished by different phases when acting on the left and right chiral fields: \(\psi_L^{CP_\chi} = e^{\mp2i\chi} i\gamma_4 C \bar\psi^T_L\). Even though all these operators are equivalent at the operator level, the approximate chiral symmetry is spontaneously broken, so that most of these \(CP_\chi\) transformations fail to leave the vacuum invariant.  The fermion phase convention is chosen to make the physically relevant \(CP_\chi\) that leaves the vacuum invariant have the standard phase choice given in Eq.~\ref{eq:std}. In Ref.~\inlinecite{us}, we discuss the phase choice for the case when the only chirally symmetric CP violating term is \({\Theta G \tilde G}\). We showed that if the chiral and CP violating parts of the action are written as 
\(
  {\cal L} \supset {d_i^\alpha O_i^\alpha}
\)
where \(i\) is flavor and \(\alpha\) is operator index, the CP violation in the theory  is proportional to:
\begin{equation}
{\bar d \bar \Theta \myRe \frac {d_i^\alpha}{d_i}} -
{|d_i| \myIm \frac{d_i^\alpha}{d_i}}\,,
\end{equation}
where
\begin{equation}
\frac1{\bar d} \equiv \sum_i \frac1{d_i}, \ \bar\Theta = \Theta - \sum_i \phi_i, \hbox{ and }
d_i \equiv |d_i| e^{i\phi_i} \equiv \frac{\sum_\alpha d_i^\alpha \langle \Omega | \myIm O_i^\alpha | \pi\rangle}
                              {\sum_\alpha \langle \Omega | \myIm O_i^\alpha | \pi\rangle}\,.
\end{equation}
Notice that CP violation depends only on {\(\bar\Theta\)} and on a {\em mismatch} of phases between \({d_i^\alpha}\) and \({d_i}\).  In what follows, we assume that \(d_i\) are dominated by the mass terms, and we choose the fermion phases such that the mass term is real.  

\subsection{Operator Basis}

In Table~\ref{tab:ops}, we enumerate the CP violating dimension 5 operators allowed by the BRST symmetry after gauge fixing to the Landau gauge (or, more generally, to any \(R_\xi\) gauge). They include both gauge invariant operators \(O\) that do not vanish by equations of motion, and gauge variant operators \(N\) that do\rlap.\footnote{At dimension 5, no CP violating operators containing the Fadeev-Popov ghosts are allowed by the BRST symmetry.}\spacefactor\sfcode`\.{} Under renormalization, their mixing structure can be written as 
\begin{equation}
  \left(\begin{array}{c}O\\N\end{array}\right)_{\rm ren} =
  \left(\begin{array}{cc}Z_O&Z_{ON}\\0&Z_N\end{array}\right)
  \left(\begin{array}{c}O\\N\end{array}\right)_{\rm bare}\,.
\end{equation}
In Ref.~\inlinecite{us}, we describe a momentum subtraction scheme, RI-$\tilde{\rm S}$MOM, that uses the \(\overline{\rm MS}\) quark masses when they appear explicitly in the operators. This scheme is defined by imposing the condition that certain projections of the truncated Green's functions of operators between quark and gluon states take on their tree-level value.  The external momenta are chosen to be symmetric, non-exceptional and to remove the non-1PI quark contributions.  The finite renormalizations that connect this scheme to the \(\overline{\rm MS}\) scheme in the continuum limit are also provided there.

\begin{table}[tp]
\begin{center}
  \begin{tabular}{|lll|}
\hline
    {\(i g\bar\psi \tilde\sigma^{\mu\nu}G_{\mu\nu}t^a\psi\)}&
    \(\partial^2 {\left(\bar\psi i\gamma_5t^a\psi\right)}\)&
    {\(\frac{ie}2 \bar\psi\tilde\sigma^{\mu\nu}F_{\mu\nu}\left\{Q,t^a\right\}\psi\)}\\
\hline
  \(\Tr\left[Mt^a\right]\partial_\mu\left({\bar\psi\gamma^\mu\gamma_5\psi}\right)\)&
  \(\frac12 \partial_\mu \left.\left(
    {\bar\psi\gamma^\mu\gamma_5\left\{M,t^a\right\}\psi}
    \right)\right|_{\rm traceless}\)&\\
  \(\Tr\left[MQ^2t^a\right]\frac12 {\tilde F_{\mu\nu}F^{\mu\nu}}\)&
  \(\Tr \left[Mt^a\right] \frac12 {\tilde G_{\mu\nu}^aG^{\mu\nu a}} \) &\\
\hline
  \(\frac12 {\bar\psi i\gamma_5\left\{M^2,t^a\right\}\psi}\)&
  \(\Tr\left[M^2\right] {\bar\psi i\gamma_5t^a\psi}\)&
  \(\Tr\left[Mt^a\right] {\bar\psi i\gamma_5M\psi}\)\\
\hline
  \(i\bar\psi_E\gamma_5t^a\psi_E\)&
  \(\myRe \partial_\mu\left[\bar\psi_E\gamma^\mu\gamma_5 t^a\psi\right]\)&\\
  \(\myRe \bar\psi\gamma_5\slashed\partial t^a\psi_E\)&
  \(\myRe \frac{ie}2 \bar\psi\left\{Q,t^a\right\}\slashed A^{(\gamma)}\gamma_5\psi_E\)&\\
\hline
  \end{tabular}
\end{center}
\caption{Flavor diagonal CP violating dimension 5 operators in the two flavor theory allowed by the BRST symmetry in Landau gauge. The mass matrix \(M\) and the charge matrix \(Q\) are assumed real and flavor diagonal, \(t^a\) stands for either an isotriplet or an isosinglet diagonal flavor generator. The subscript `traceless' indicates that the flavor trace is subtracted from the anti-commutator.  We use the notation \(\psi_E \equiv (i\slashed D - m) \psi\) for a fermion field that is zero by the equations of motion.}
\label{tab:ops}
\end{table}


\subsection{Form Factors}

The electric dipole moment can be calculated from the matrix element of the electromagnetic current. In fact, it is one of the zero-momentum electromagnetic form factors.  For spin 1/2 particles like the neutron \(N\), the interaction of the electromagnetic current \(V_\mu(q)\) is given by the Dirac \(F_1\) and Pauli \(F_2\) form factors, the electric dipole form factor \(F_3\) and the anapole form factor \(F_A\):
\begin{eqnarray}
\langle N | V_\mu(q) | N \rangle & = &
   \overline {u}_N \left[ 
         \gamma_\mu\;F_1(q^2) + i \frac{[\gamma_\mu,\gamma_\nu]}2 q_\nu\; \frac{F_2(q^2)}{2 m_N} \right.\nonumber\\
 &&\qquad
         \left.
               {} + {(2 i\,m_N \gamma_5 q_\mu - \gamma_\mu \gamma_5 q^2)\;\frac{F_A(q^2)}{m_N^2}}
         {} + {\frac{[\gamma_\mu, \gamma_\nu]}2 q_\nu \gamma_5\;\frac{F_3(q^2)}{2 m_N}} \right] u_N\,,
\end{eqnarray}
where \(u_N\) represents the free neutron spinor and \(m_N\) is the neutron mass.  The Sachs electric and magnetic form factors are defined in terms of these as \(G_E=F_1 - (q^2/4M^2) F_2\) and  \(G_M=F_1 + F_2\) respectively.  The zero momentum limit of these form factors give the charges and dipole moments: thus, for the neutron, we have the electric charge is \(G_E(0) = F_1(0) = 0\), the (anomalous) magnetic dipole moment is \(G_M(0)/2 M_N = F_2(0) / 2 M_N\), and the electric dipole moment is \(F_3(0)/2 m_N\). The form factors \(F_A\) and \(F_3\) violate parity P, and \(F_3\) violates CP as well.  

\begin{figure}[tp]
\begin{eqnarray*}
&\vcenter{\hbox{\includegraphics[width=0.12\textwidth]{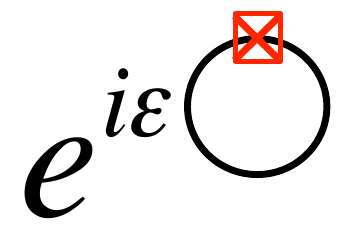}}} \times {}&\\[3\jot]
&\left(\;\vcenter{\hbox{\includegraphics[width=0.32\textwidth]{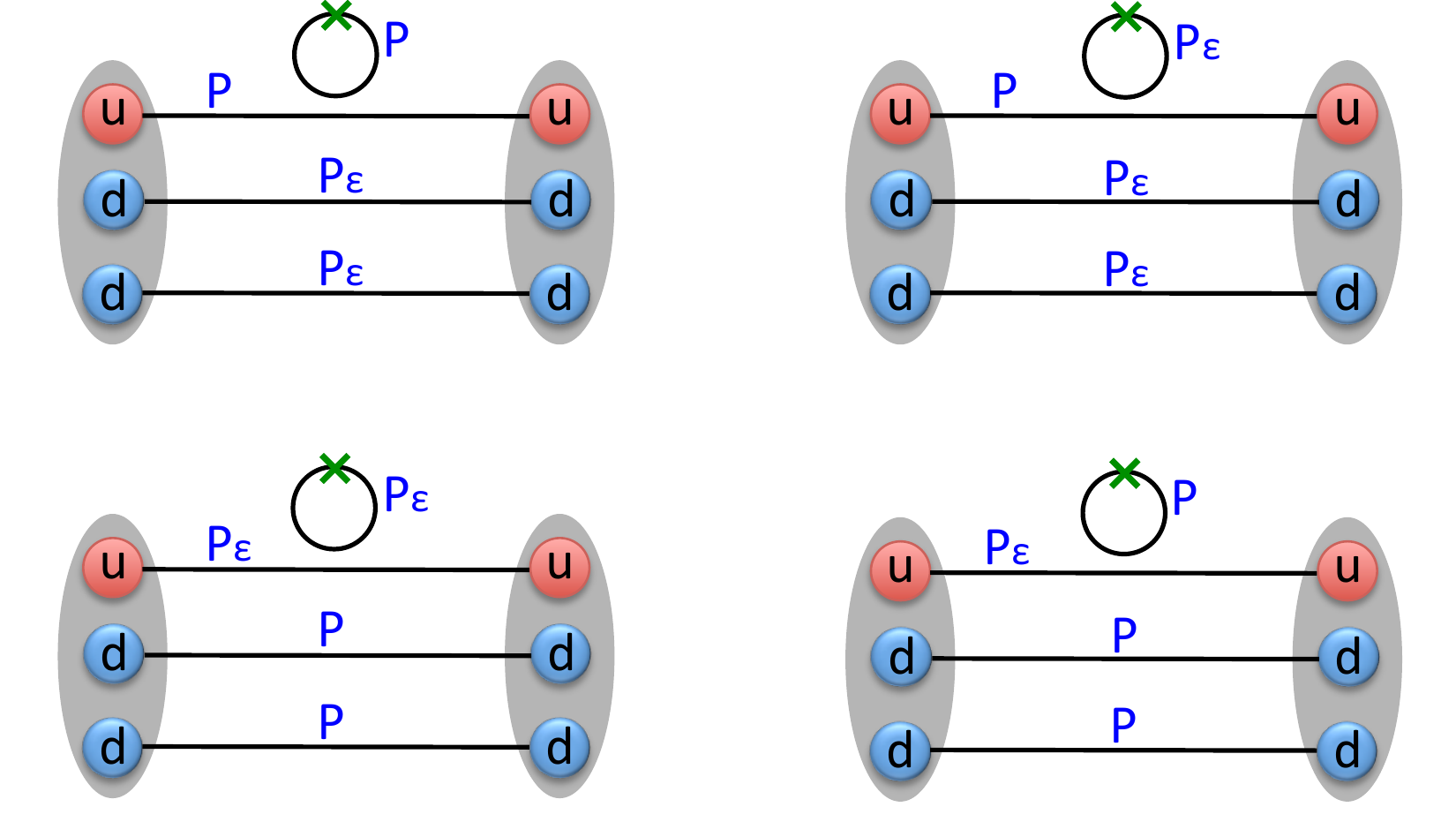}}} + 
 \vcenter{\hbox{\includegraphics[width=0.32\textwidth]{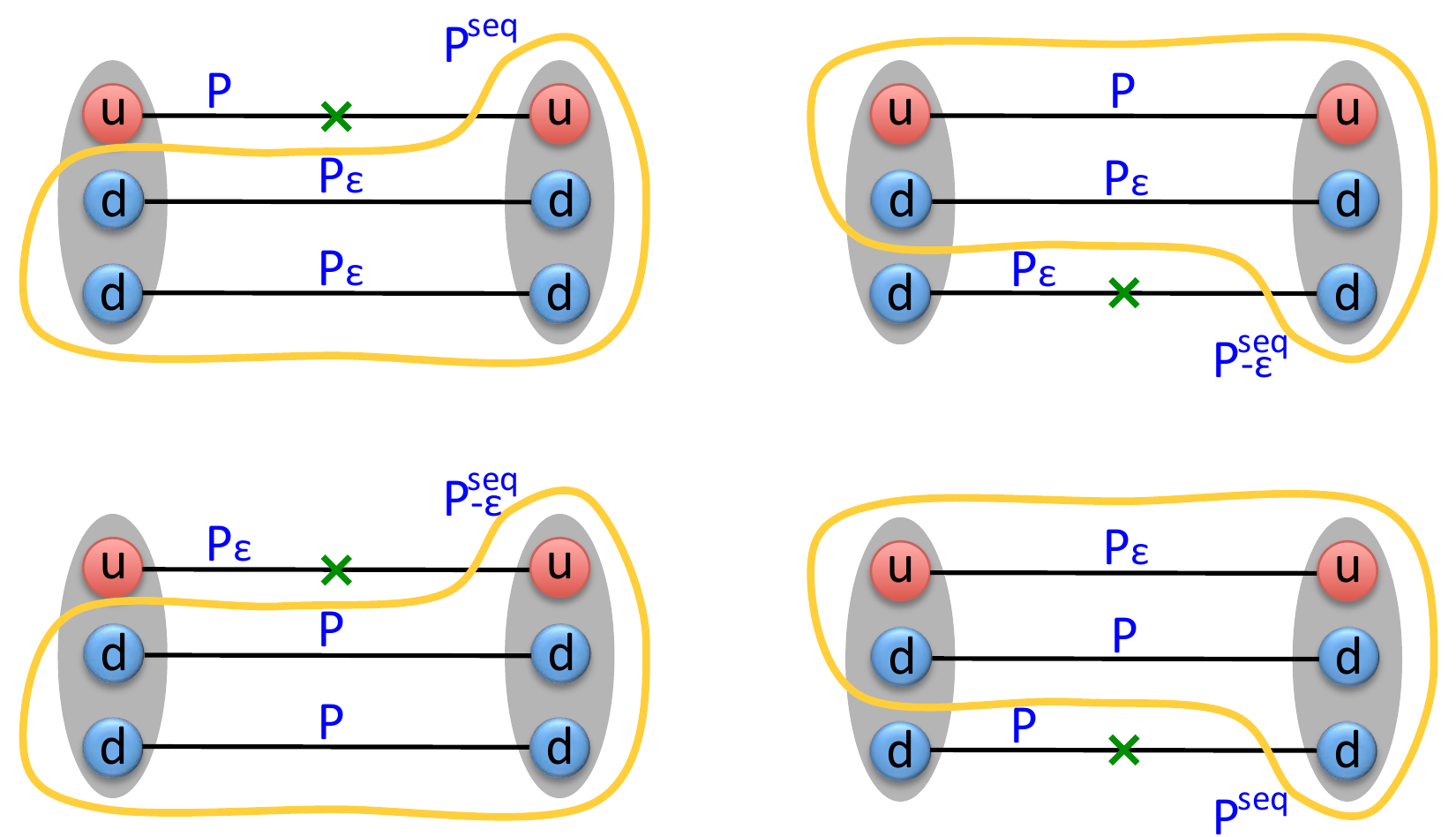}}}\;\right)&
\end{eqnarray*}
\caption{The calculation of the electromagnetic form factors in the presence of the chromo-electric operator using the Schwinger source method. The calculation proceeds by tying together propagators with (marked with \(\epsilon\) or \(-\epsilon\)) and without the addition of the chromo-electric operator. The superscript `seq' represents a sequential propagator. The green cross stands for the electromagnetic current and the red cross in a square, the chromo-electric operator.}
\label{fig:EDM}
\end{figure}

\subsection{Projection}

We evaluate these form factors on the lattice by measuring the three point correlator of the electromagnetic current with the neutron operator
\(
    N \equiv {\bar d}^c \gamma_5 \frac{1+\gamma_4}2 u\; d
\)
in the vacuum state \(|\Omega\rangle\). This can be expanded in a complete basis of states \(|N\rangle, |N'\rangle, \ldots\) to which the neutron operator couples:
\begin{equation}
\langle \Omega | {N(\vec0,0) V_\mu(\vec q,t) N^\dagger(\vec p,T)} | \Omega \rangle =
 \sum_{N,N'} u_N e^{-m_N t}\; {\langle N | V_\mu(q) | N' \rangle}\; e^{-E_{N'} (T-t)}{\overline u}_N\,.
\end{equation}%
For simplicity, we project on to a single component of the neutron spinor using the projector
\(
{\cal P} = \frac12 (1+\gamma_4) (1+i\gamma_5\gamma_3)
\).
Noting that in  presence of CP violation, the free neutron spinor satisfies
\(
u_N {\overline u}_N = {e^{i\alpha_N \gamma_5}} (i\slashed p + m_N) {e^{i\alpha_N \gamma_5}}
\)
for some CP violating phase angle \(\alpha_N\), and assuming that the neutron operator couples primarily to a single state \(N'=N\), we can extract the form factors from
\begin{eqnarray}
\Tr {\cal P} \langle \Omega | {N V_3 N^\dagger} | \Omega \rangle &\propto& i m_N q_3 G_E
 - 2 i\, (q^2_1 + q^2_2) {F_A} - \frac{q_3^2}2 {F_3} \nonumber\\
 &&\qquad {}+
{\alpha_N} m_N (E_N-m_N) F_1 
+ {\alpha_N} [ m_N (E_N - m_N) + \frac {q_3^2}2] F_2
\label{eq:F3}
\end{eqnarray}

\section{Lattice Calculation}

The calculation of the electromagnetic form factors in the presence of a chromo-EDM operator na\"{\i}vely needs the evaluation of a four point function.  The technology for such calculations is at its infancy.  To avoid this problem, we choose to follow the Schwinger source method, in which the chromo-EDM operator is added to the Lagrangian with a coefficient, \(\epsilon\).  Derivatives with respect to \(\epsilon\) of any matrix element calculated with this Lagrangian and evaluated at \(\epsilon=0\) then `inserts' the chromo-EDM operator.

\subsection{Schwinger source method}

Since the quark chromo-EDM operator is a quark bilinear, the addition of this operator to the Lagrangian can be thought of as an addition to the Dirac operator. This change can be implemented by changing the fermion propagator
{\begin{equation}
   \left(\Dslash + m - \frac r2 D^2 + c_{sw} \Sigma^{\mu\nu} G_{\mu\nu}\right)^{-1} \mathbin{{\longrightarrow}}
   \left(\Dslash + m - \frac r2 D^2 + \Sigma^{\mu\nu} ( c_{sw} G_{\mu\nu} + i \epsilon {\tilde G_{\mu\nu}} )\right)^{-1}\!\!\!\!
\end{equation}}%
and multiplying the fermion determinant by the `reweighting factor'
{\begin{eqnarray}
\frac {\det ( \Dslash + m - \frac r2 D^2 + \Sigma^{\mu\nu} ( c_{sw} G_{\mu\nu} + i \epsilon {\tilde G_{\mu\nu}} )}
          {\det ( \Dslash + m - \frac r2 D^2 + c_{sw} \Sigma^{\mu\nu} G_{\mu\nu} )}\span\omit\span\nonumber\\
\qquad\qquad\qquad\qquad&=&
\exp \Tr \ln \left[1 + i \epsilon\, {\Sigma^{\mu\nu} \tilde G_{\mu\nu}} ( \Dslash + m - \frac r2 D^2 + c_{sw} \Sigma^{\mu\nu} G_{\mu\nu} )^{-1}\right]
\nonumber\\
&\approx&
\exp \left[ i \epsilon \Tr {\Sigma^{\mu\nu} \tilde G_{\mu\nu}} ( \Dslash + m - \frac r2 D^2 + c_{sw} \Sigma^{\mu\nu} G_{\mu\nu} )^{-1}\right]\,.
\end{eqnarray}}%
Schematically, the entire calculation of the electromagnetic form-factors in the presence of \(u\) and \(d\) quark chromo-electric dipole moments is illustrated in Fig.~\ref{fig:EDM}.

\begin{table}[tp]
\begin{center}
  \begin{tabular}{|c|c|c|}
    \hline
    Accuracy & $\epsilon=0.005$ & $\epsilon=0.01$ \\
    \hline
    $10^{-8}$ & 85\% & 86\% \\
    $10^{-3}$ & 51\% & 66\% \\
    $5 \times 10^{-3}$ & 28\% & 45\% \\
    \hline
  \end{tabular}
\end{center}
\caption{Cost of inversion of the modified Dirac operator compared to that of the unmodified operator.  The details of the ensemble and method are given in the text.}
\label{tab:cost}
\end{table}

\subsection{Propagator inversion}

We studied the inversion of the modified Dirac operator using clover valence quarks on \(a\approx 0.12\hbox{fm}\), \(m_\pi \approx 310\hbox{MeV}\) HISQ ensembles from the MILC collaboration~\cite{MILC}. A single application of the modified Dirac operator was only 7\% more expensive once the chromo-electric field has been precalculated, and we observed that the condition number of the modified Dirac operator was within 5\% of the unmodified Dirac operator. Using the BiCGStab algorithm implemented in Chroma software suite~\cite{Chroma} and using the \(\epsilon=0\) solution as an initial guess, the extra inversion cost of the modified operator was 28--86\% of the cost of the inversion of the unmodified operator, as shown in Table~\ref{tab:cost}.  Overall, the calculation of connected electromagnetic current measurement on each configuration is only about 50\% larger than the cost of the same measurements in the absence of the chromoelectric operator.

\section{Numerical Tests}

\begin{figure}[tp]
\begin{minipage}[t]{0.48\textwidth}
\begin{center}
{%
\includegraphics[width=0.48\textwidth]{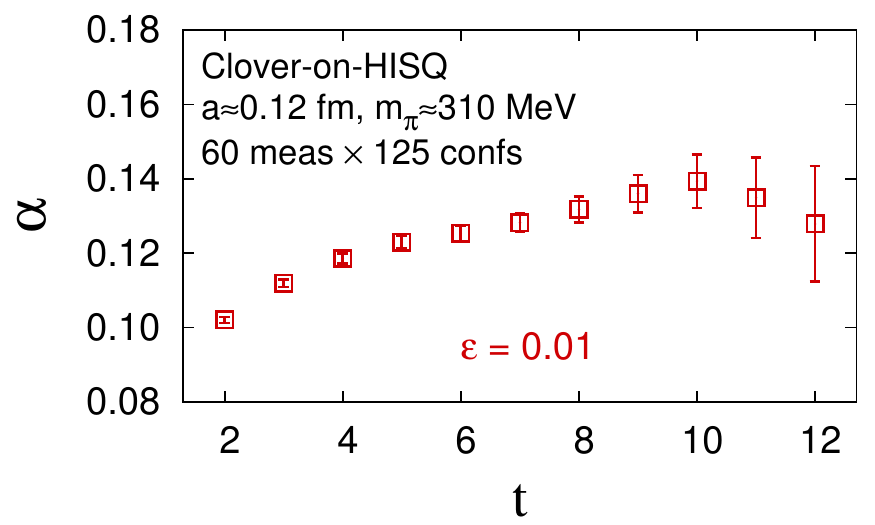}%
\includegraphics[width=0.48\textwidth]{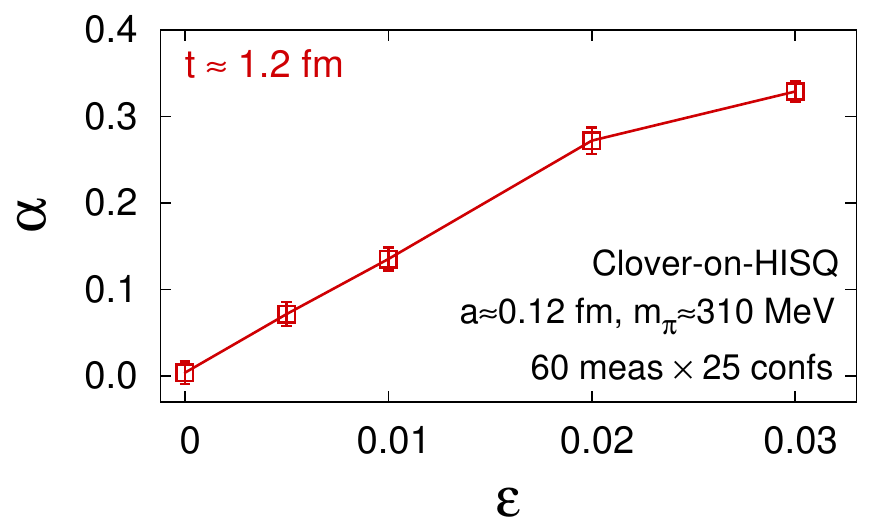}%
}
\end{center}
\caption{The phase \(\alpha\) of the connected two point function of the neutron due to a chromo-electric moment for the down quark as a function of the source-sink separation \(t\) (left) and the strength of the quark moment \(\epsilon\) (right).}
\label{fig:alpha}
\end{minipage}\hspace{0.01\textwidth}
\begin{minipage}[t]{0.48\textwidth}
\begin{center}
\includegraphics[width=0.48\textwidth]{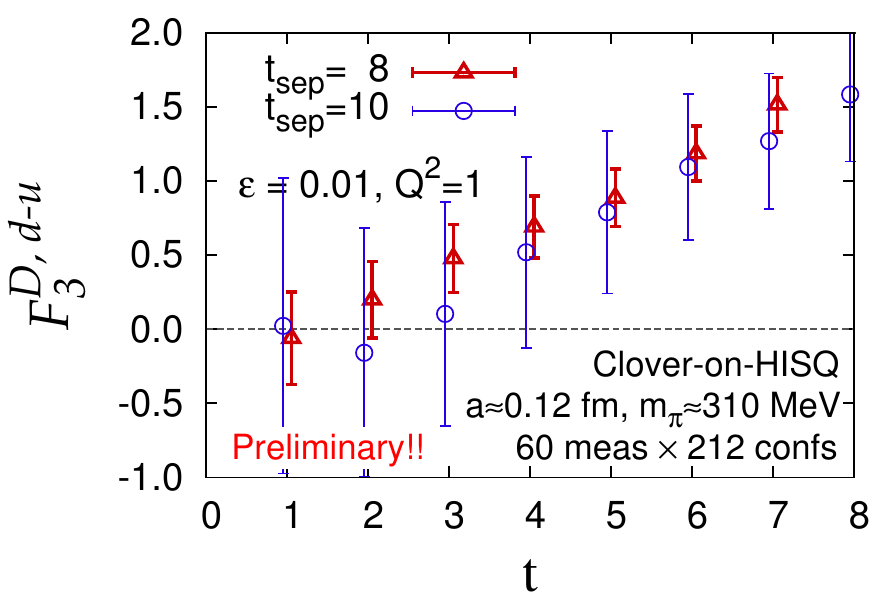}%
\includegraphics[width=0.48\textwidth]{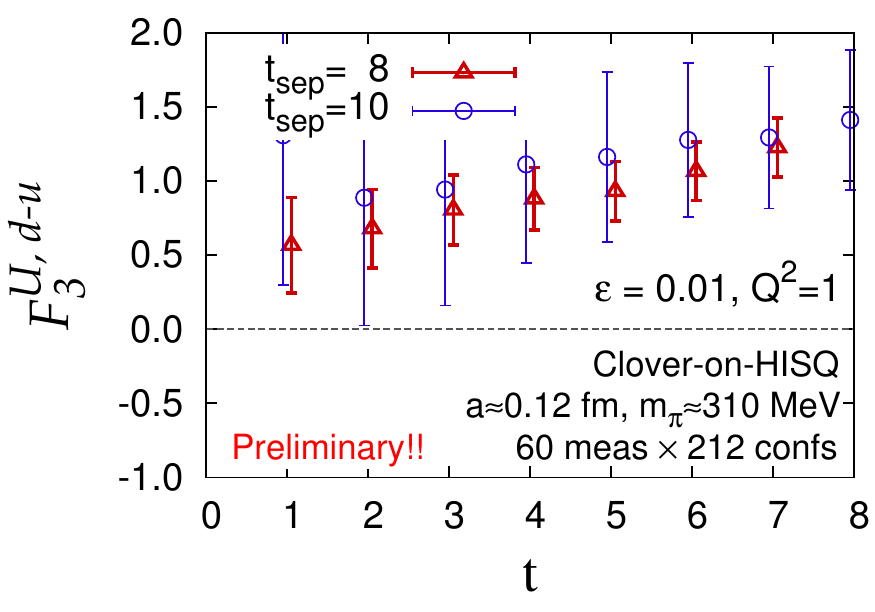}%
\end{center}
\caption{Signal in the connected isovector \(F_3\) form factor of the neutron due to a chromo-electric dipole moment of the up (left) and down (right) quarks as a function of the operator insertion time.  \(t_{\rm sep}\) indicates the source-sink separation in lattice units.}
\label{fig:F3}
\end{minipage}
\end{figure}

The Schwinger source method relies on taking the derivative of the matrix element with respect to \(\epsilon\) at \(\epsilon=0\).  The addition of the higher dimension operator to the Lagrangian, however, makes the theory nonrenormalizable at finite values of \(\epsilon\).  In other words, one needs to keep \(\epsilon\) large enough so that the differences used to evaluate the derivatives are not dominated by noise, and yet small enough, \(\epsilon \lesssim 4 \pi a \Lambda_{\rm QCD}\), so that the \(O(a^{-1})\) divergences are under control. In Fig.~\ref{fig:alpha}, we show that the parameter \(\alpha_N\) can be calculated from the connected nucleon two point function, and is linear in \(\epsilon\).\looseness-1

In Fig.~\ref{fig:F3}, we show the signal in the connected diagrams of the \(F_3\) form factor obtained using Eq.~\ref{eq:F3} and the determination of \(\alpha_N\).  The signal is non-zero, but a plateau is not yet visible in the preliminary data.

\section{Conclusions}

An extraction of \(F_3\) and neutron electric dipole moment will need control over statistics, excited state effects and operator mixing. Also, the continuum limit of this matrix element has an \(O(a^{-2})\) divergent mixing with lower dimensional operator that need to be subtracted non-perturbatively. In discretizations like our mixed action formalism without chiral symmetry, there are additional divergences that also need to be controlled.  Calculations are currently underway.

\end{document}